\documentclass[11pt,twoside]{article}
\usepackage[dvips]{graphics}
\usepackage{newpasp}
\begin{document}

\title{Could merged star-clusters build up a small galaxy?}

\author{Michael Fellhauer}
\affil{ARI Heidelberg, Germany, mike@ari.uni-heidelberg.de}
\author{Pavel Kroupa}
\affil{ITA \& MPIA Heidelberg, Germany, pavel@ita.uni-heidelberg.de}

\begin{abstract}
  We  investigate the behaviour of a cluster of young massive star
  clusters (hereafter super-cluster)  in the tidal field of a host
  galaxy with a high-resolution particle-mesh code, {\em Superbox}.   
  Specifically we want to establish if and how such super
  star-clusters merge and carry out a detailed study of the
  resulting merger-object.  This merger-object shows either the
  properties of a compact spherical object or the elongated
  (`fluffy') shape of dSph-galaxies depending 
  on the initial concentration of the super-cluster.
\end{abstract}
\vspace*{-1.0cm}

\section{Introduction}
\label{sec:intro}

Interacting galaxies like the Antennae (NGC~4038/4039; Whitmore \&
Schweizer 1995) or Stephan's Quintet (HCG~92; Hunsberger 1997) show
much star-burst activity in their tidal features.  High resolution
images from the HST resolve these regions into many compact groups of
young massive star clusters (i.e. super-clusters) and/or tidal-tail
dwarf galaxies, with typical radii of 100--500~pc.  Here we aim to
study the future fate of these super-clusters. \\ We begin with an
overview of the numerical method and explain the setup of our
simulations.  We then show results obtained so far and conclude with
an outlook on future work we intend to pursue.

\section{Superbox}
\label{sec:superbox}

{\em Superbox} is a hierarchical particle-mesh code with 
high-resolution sub-grids focusing on the cores and the
star-clusters as a whole, and moving with them through the
simulation area (Fellhauer et al. 2000). The code has, for
particle-mesh codes, a highly-accurate
force-calculation based on a nearest grid-point (NGP) scheme.
The main advantages of {\em Superbox} are it's speed and the
low memory requirement which makes it possible to use a high
particle number with high grid-resolution on normal desktop
computers.  

\section{Setup}
\label{sec:setup}

As a model for our massive star-clusters we use, for each,
Plummer-spheres with 100,000 particles, a Plummer-radius 
$R_{\rm pl} = 6$~pc and a cutoff radius $R_{\rm cut} = 15$~pc,
giving  a total mass of $10^{6}\ {\rm M}_{\odot}$ and crossing
time of $1.4$~Myrs.  Twenty of these clusters are placed in a
compact group  orbiting in a logarithmic potential of the parent
galaxy, 
\begin{eqnarray}
  \label{eq:potential}
  \Phi & = & \frac{1}{2} \, v_{\rm circ}^{2} \, \ln \left( R_{\rm
      gal}^{2} + r^{2} \right),
\end{eqnarray}
with $R_{\rm gal} = 4$~kpc and $v_{\rm circ} = 220$~km/s. The
case $v_{\rm circ}=0$ is dealt with in Kroupa (1998).  The
distribution of the super-cluster is also Plummer-like with
different Plummer-radii (Table~1).  The tidal radius is $\approx$
2.4~kpc at  apo-galacticon and 1.2 kpc at peri-galacticon.  The
orbits have the same eccentricity in all cases, and begin at
apo-galacticon ($x = 60$~kpc, $y,z = 0$) with $v_{y} = 150$~km/s
($v_{x}=v_{z}=0$). 
\vspace*{0cm}
\begin{table}[h!]
  \begin{center}
    \caption{Radial scale-length and crossing time for the
      different runs} 
    \label{tab:scalelength}
    \begin{tabular}[h!]{c|r|r}
      simulation & scale-length & $T_{\rm cr}^{\rm sc}$ \\
      \tableline 
      run6 & 300 pc & 108.4 Myr \\
      run5 & 150 pc &  38.3 Myr \\
      run7 &  75 pc &  13.5 Myr \\ \tableline \tableline
    \end{tabular}
    \label{tab:runs}
  \end{center}
\end{table}
\vspace*{-0.5cm}
\section{First Results}
\label{sec:results}

\subsection{Global properties of the Super-Cluster}
\label{sec:globalresults}

In all runs some of the clusters merge very rapidly within the
first $100$~Myrs as seen in Fig.~1.
\begin{figure}[h!]
  \begin{center}
    \plotfiddle{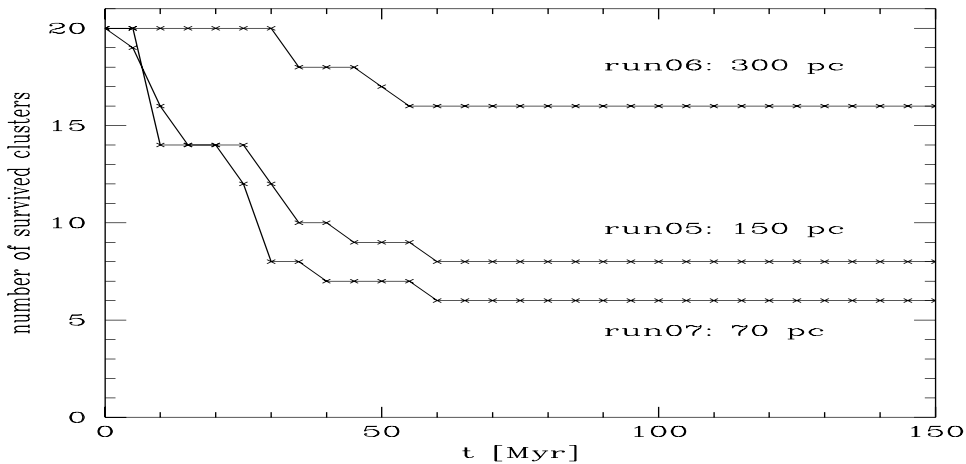}{5cm}{0}{65}{32}{-200}{-60}
    \caption{Number of surviving clusters vs.\ time. The clusters
      are taken as merged to the merger-object when the 80~per cent
      mass-shell starts to expand and the cluster is known to end
      in the merger-object.} 
    \label{fig:mergetime}
  \end{center}
\end{figure}
The number of surviving clusters drops with increasing
concentration of the super-cluster (SC).  The orbits of the
clusters inside their super-cluster also change rapidly due to
the very short relaxation-time of the SC ($T_{\rm relax} \approx
T_{\rm cr}^{\rm sc}$).  Thereafter the merger-object and the
surviving clusters move on epicycles around the orbit of the
original SC about the galactic centre, with an increasing
amplitude and period due to tidal heating (Fig~2).  The surviving
star-clusters remain in the vicinity of the merger-object
(Fig.~3). 
\begin{figure}[t!]
  \begin{center}
    \plotfiddle{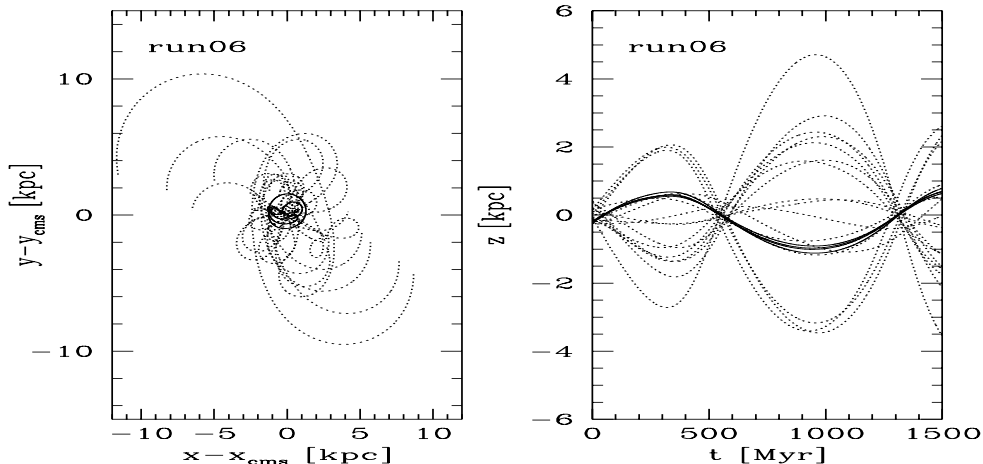}{5cm}{0}{65}{32}{-200}{-60}
    \caption{Left: internal motion of the merger-object and the
      surviving star-clusters in the $x$-$y$-plane; right:
      $z$-amplitude vs.\ time; internal orbits are soon lost and
      epicycles with increasing amplitude and period are evident;
      solid lines: merged clusters, dotted lines: survived
      clusters} 
    \label{fig:xycmszheight}
  \end{center}
\end{figure}
In the intermediate concentration case (run5) we find 2
merger-objects.  To check whether this was just a chance event, a
second run with a different random number seed for the
star-cluster positions also gave 2 objects.  Interesting in this
context is that Theis (1996) found that two clusters form in a
cold collapse of a stellar system in an external tidal
field. Violent relaxation in an external tidal field needs
further addressing in the present context. 

\subsection{Internal properties of merger-objects}
\label{sec:internalresults}

\begin{figure}[t!]
  \begin{center}
    \plotfiddle{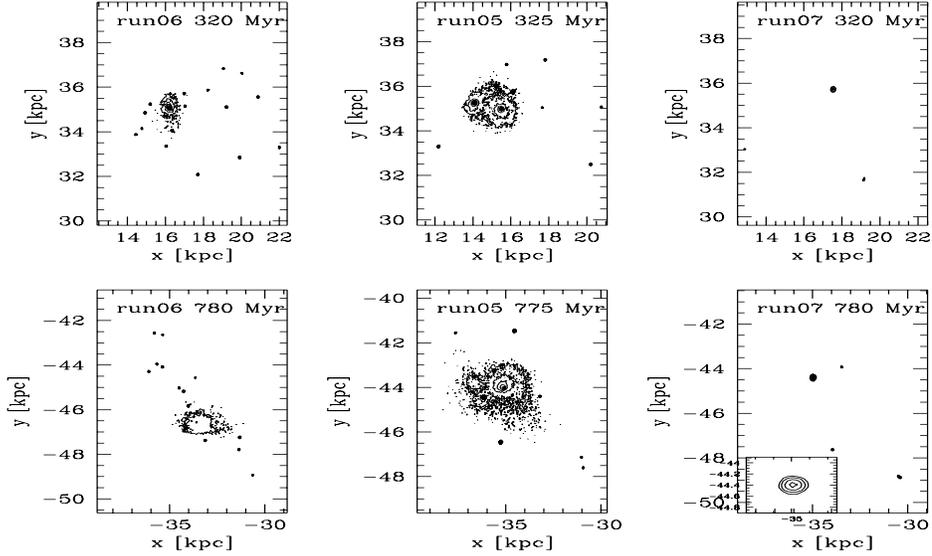}{7cm}{0}{65}{40}{-200}{-60}
    \caption{Contour-plots of the different simulations after 320 
    (upper panels) and 780 (lower panels) Myrs; the lower right
    panel also shows a blow-up of the spherical merger-object} 
    \label{fig:contour}
  \end{center}
\end{figure}
In those cases where the SC has a low concentration initially,
the resulting merger-system is an extended object (several kpc)
with a low density and an off-centre nucleus (Fig.~3).  The
radial density profile follows an exponential distribution
(Fig.~4). The velocity-dispersion in these objects drops to
$\approx$ 5~km/s (Fig.~5), which is comparable with the measured
dispersions of dSph-galaxies in the Milky Way, and is slightly
anisotropic.
\begin{figure}[h!]
  \begin{center}
    \plotfiddle{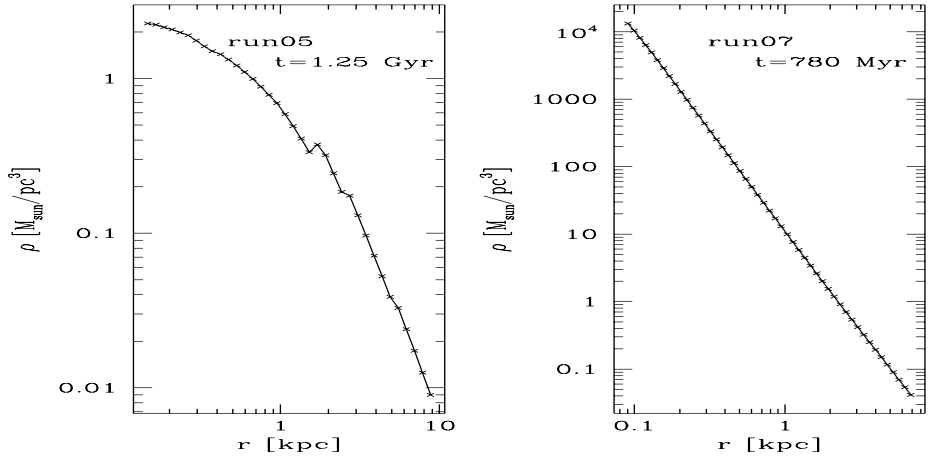}{5cm}{0}{65}{32}{-200}{-60}
    \caption{Mass density profile of the
      merger-object for a low concentration (left) and a
      high concentration run (right)}  
    \label{fig:density}
  \end{center}
\end{figure}
\begin{figure}[h!]
  \begin{center}
    \plotfiddle{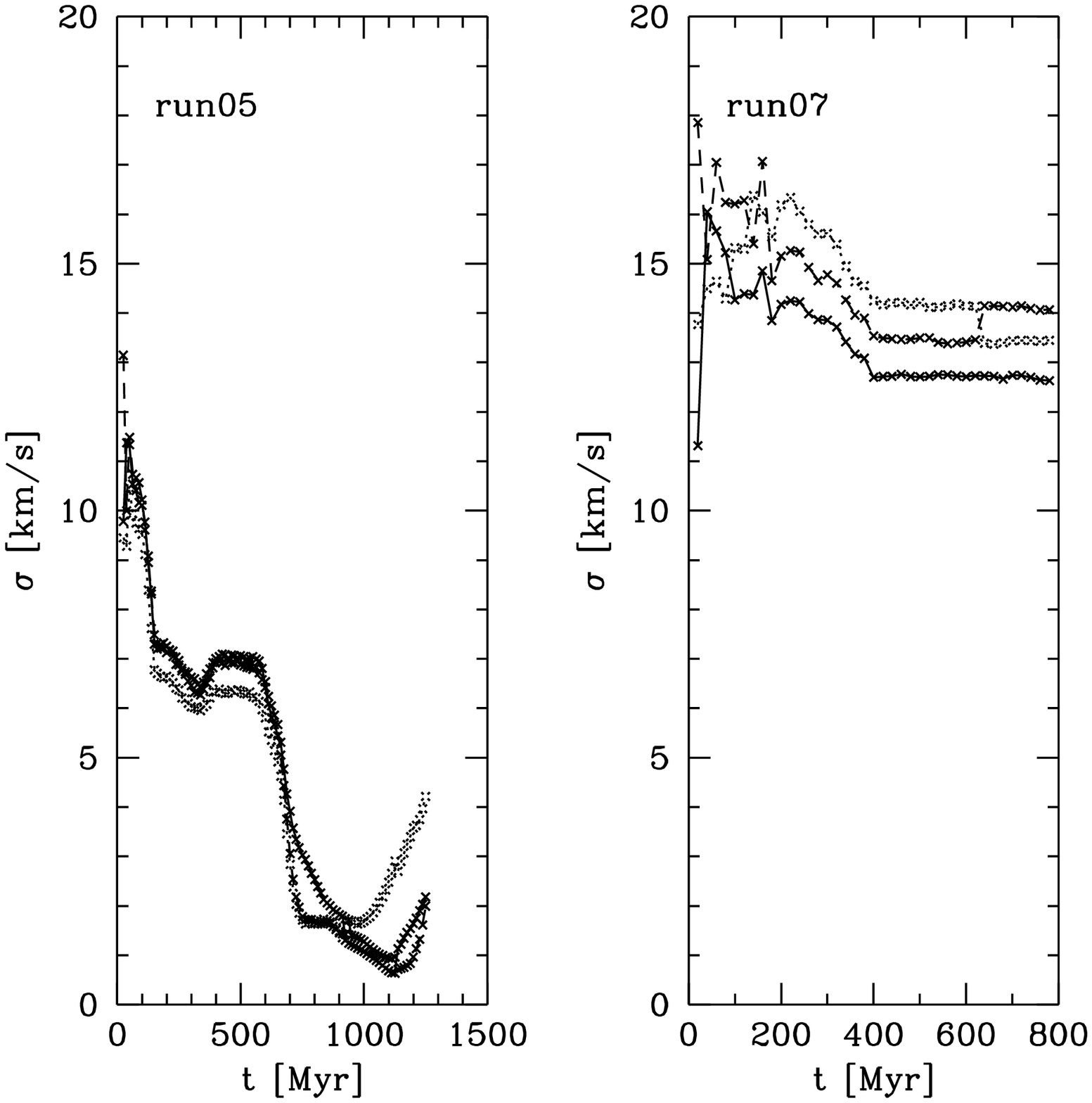}{5cm}{0}{65}{32}{-200}{-60}
    \caption{Internal velocity dispersion of the merger-object
      measured out to the 5\%-mass-shell of a low concentration
      (left) and a high concentration run (right)} 
    \label{fig:velocity}
  \end{center}
\end{figure}
In the case of the most concentrated SC, the merger-object is a
dense compact spheroidal object (Fig.~3) with a high density core
($\approx 10^{4}\ {\rm M}_{\odot}/{\rm pc}^{3}$).  The density
profile follows a power-law with $r^{-3}$ out to 0.8-1.0~kpc
(Fig.~4).  The velocity-dispersion of this spheroidal dwarf
galaxy is anisotropic and around 15~km/s (Fig.~5).  It has a
half-mass radius of about 300~pc.  The size of this object is too
large in comparison with even the biggest of the globular
clusters, and the central density is far too high to be
comparable with any dE-, dIrr- or dSph-galaxy.  However, it has
properties similar to the initial satellites studied by Kroupa
(1997), and is thus likely to evolve to a dSph-like
satellite. But further investigation of this issue is necessary. 

\section{Conclusion and Outlook}
\label{final}

We found that even if new high resolution images show that most
of the so-called tidal dwarf galaxies are clusters of young
compact massive star clusters they are likely to merge within a
short time-scale.  The properties of the merger-objects differ
with the scale-length of their initial distribution.  We found
large fluffy objects with similar properties as the local
dSph-galaxies as well as very compact and massive spheroidal
objects, which, however, may be similar to the progenitors of
some of the local present-day dSph satellites. \\ 
In the course of future work we intend to investigate the
influence of the choice of orbit around the parent galaxy and how
this alters the results.  We will focus on the transition between
bound and unbound objects, and look for a region in the space of
parameters where 2 merger-objects (binary system) are more likely
to form.  Our further research will also address the future fate
of the merger-objects and their possible counterparts in reality.


\acknowledgements
We thank C. M. Boily for providing helpful comments, and
R. Spurzem for supporting this project at the ARI.

\vspace{0.5cm} 
{\bf U. Fritze von Alvesleben}: It would be interesting to try
and see if the age distribution among clusters that are clustered
is different (younger) from that of YSCs distributed more
homogeneously.  What do you expect to happen with the merged
clusters that you compared to a dSph.  Isn't it bound to sink
into the core of the merger remnant by dynamical friction ?  The
bulk of the YSCs is at $d \leq 10$~kpc from the nucleus of
NGC~4038, i.e. not further away than NGC~4039 (nucleus)! \\  
{\bf Answer}: We assume that at least some of the tidal-tail
dwarfs seen to form in outer ($>30$~kpc) tidal arms are composed
of clusters of young massive star clusters.  For our models we
use an analytic galactic potential, because in the mass range of
the merger-object of about $10^{7}$~M$_{\odot}$, dynamical
friction does not play a significant role. \\
{\bf E. Grebel}: Could you comment on how the merged clusters
will resemble a dSph galaxy in their properties (E.g. dSph don't
show rotation, have very low density and surface brightness,
etc.) ? \\  
{\bf Answer}: It is too early to quantify the reply in detail,
but we expect the merged object to show properties that could
make it look similar to the progenitors of some of the dSph
satellites.  The merged object is spheroidal, and has a high
specific frequency of globular clusters, and low angular
momentum, which however depends on the initial conditions. It's
stellar population contains stars from the mother galaxy, as well
as stars formed during the star burst, and maybe stars (and
clusters) formed during a possible later accretion event of a
co-moving gas cloud. Many of these issues are discussed in Kroupa
(1998).\\ 
{\bf J. Gallagher} (comment): Since super-star-clusters are often
born in groups -- the luminous clumps -- destruction via cluster
merging is of general interest.  For example, if this process
reduces the survival rate of massive-star-clusters, it might help
to explain why intermediate age examples seem to be rare. \\ 
{\bf D. McLaughlin}: What evidence is there that the clustered
clusters in the Antennae will actually merge ?  There are $\approx
10^{9}$~M$_{\odot}$ clouds of gas in this system, so it may be that
young clusters are clustered because several form in any given cloud,
but they disperse after gas-loss. With no information on either the
cluster-cluster velocity dispersion or the time when the parent cloud
was disrupted, it would be difficult to rule out this possibility. \\
{\bf Answer}: This is an important issue, and very similar to
the problem of forming bound star clusters. While not disproving
rapid dispersal entirely, the argument which makes rapid
dispersal less likely is as follows: The cluster-cluster velocity
dispersion, $\sigma_{\rm clcl}$ is either small, which will lead
to a bound merger object. If $\sigma_{\rm clcl}$ is near to
virial for the {\it stellar} mass in the super-cluster, then our
models take care of that. If, however, $\sigma_{\rm clcl}$ is
virial for the stars and a much larger mass in gas, then
$\sigma_{\rm clcl}>20$~km/s assuming a star-formation efficiency
of 20~per cent and a pre-gas removal super-cluster configuration
as in run5 here. Thus, within 10~Myr, the object will have
expanded to a radius of at least 350~pc. Since many of the
super-clusters are still very concentrated, and about 10~Myr old, 
rapid expansion does not appear to be taking place, especially so
since there are at least about 10~observed super clusters and they
would have had to start in unrealistically concentrated configurations
for them to appear with the sizes they have now.  This is further
discussed in (Kroupa 1998).


\end{document}